\documentstyle[amstex,12pt]{article}                 
\newlength{\dinwidth}
\newlength{\dinmargin}
\newcommand{\CC}{{\mbox{{\small $ \Bbb C$}}}}
\newcommand{\RR}{{\mbox{{\small $ \Bbb R \,$}}}}
\newcommand{\Wb}{{\cal W}_\beta}
\newcommand{\Tb}{{\cal T}_\beta}
\newcommand{\Tr}{{\mbox{Tr}}}
\begin{document}
\title{Relativistic KMS--condition and K\"all\'en--Lehmann type
representations of thermal propagators}
\author{Jacques Bros$^a$ \,  and  \, Detlev Buchholz$^b$
\\[5mm]
${}^a \,$ Service de Physique Th\'eorique, CEA--Saclay\\
F--91191 Gif--sur--Yvette, France\\[2mm]
${}^b \,$ II.\ Institut f\"ur Theoretische Physik,
Universit\"at Hamburg\\
D--22761 Hamburg, Germany}
\date{}
\maketitle
%
%

\begin{abstract} A relativistic version of the Kubo--Martin--Schwinger
boundary condition is presented which fixes the properties of
thermal equilibrium states with respect to arbitrary space--time
translations. This novel condition is a natural generalization of
the relativistic spectrum condition in the vacuum theory and has
similar consequences. In combination with the condition of locality
it gives rise to a K\"all\'en--Lehmann type representation of thermal
propagators with specific regularity properties. Possible
applications of the results and some open problems are outlined.
\end{abstract}

The theory of thermal equilibrium states in relativistic quantum
field theory has not yet reached a level of understanding which is
comparable to that of the vacuum theory. For example, there does
not yet exist a perturbation theory for thermal equilibrium states
which has proved to be consistent to any given order. Many
perturbative computations are based on ad hoc regularization
procedures whose theoretical basis is frequently not clear. As
a matter of fact, it is sometimes not even clear whether these
procedures are compatible with basic principles of relativistic
quantum field theory, such as locality.

In view of this situation it seems worthwhile to supplement the
model studies by a general analysis of thermal equilibrium states
which is based on the physical foundations of the theory. One may
hope that this complementary approach will lead to a clarification
of the general structure of thermal equilibrium states and thereby
help to set up a consistent framework for their actual computation.
It is the aim of this contribution to report on progress in this
direction \cite{BrBu1, BrBu2, BrBu3} and to indicate some
interesting open problems.

Let us begin by recalling the general principles of relativistic
quantum field theory for the case of thermal equilibrium states.
There exist various competing formulations of the theory which are
essentially equivalent. We adhere here to the real time
(Minkowski space) approach and comment later on the others.

The central object of interest are the quantum fields
$\phi (x)$
etc.\ which are labelled by the spacetime points
$x  \in \RR^4$
and possibly by further tensor or spinor indices, indicating their
transformation properties under Lorentz and internal symmetry
transformations. In order to keep the notation simple we consider
here only the case of a hermitean scalar field. We will also omit the
smoothing of the fields
$\phi (x)$
with test functions which is necessary to convert these
entities into well--defined operators.

The fundamental constraint imposed on the quantum fields in a
relativistic theory is the condition of locality (spacelike
commutativity), which reads for the case at hand
\begin{equation}
\phi(x) \phi (y) = \phi (y) \phi (x) \quad \mbox{if}
\quad (x-y)^2 < 0.
\end{equation}
This relation is an operator equality which does not depend on the
physical states which one considers. But it imposes strong constraints
on the structure of these states, as we shall see.

The thermal equilibrium states of interest here
fix a Lorentz system with
privileged coordinates $x = (x_0 , {\bf x})$.
They give
rise to correlation functions which can formally be presented
according to
\begin{equation}
\langle \phi (x_1) \cdots \phi (x_n) \rangle_\beta =
\lim \, {1 \over Z} \, \Tr \,
e^{- \beta H} \phi (x_1) \cdots \phi (x_n).
\end{equation}
Here
$\beta$
is the inverse temperature, $H$ is the Hamiltonian in the vacuum
Hilbert space, $Z$ is the partition function and
the symbol ``lim'' indicates the
fact that one first has to consider the theory in a finite box
and then to proceed to the thermodynamic limit. A rigorous version
of relation (2) has been established in a C$^*$--algebraic setting in
\cite{BuJu}.

It is a well--known fact that the correlation functions (2) of
thermal equilibrium states satisfy the Kubo--Martin--Schwinger
boundary condition \cite{HaHuWi}, cf.\ below. As a matter of fact,
this property is even the distinctive feature of thermal
equilibrium states \cite{PuWo, HaKaTr}. But, as it has been noticed
in \cite{BrBu1}, relation (2) suggests a much stronger version of the
KMS--condition in relativistic theories.

To outline the argument leading to this conclusion let us consider
the expectation value
\begin{eqnarray}
& \langle
\phi (x_1) \cdots \phi (x_m)
\phi (y_1 + z) \cdots \phi (y_n + z)
\rangle_\beta = & \\
& = \lim \, {1 \over Z} \, \Tr \,
e^{ -(\beta + i z_0)H + i {\bf z P}}
\phi (x_1) \cdots \phi (x_m)
e^{ i z_0H - i {\bf z P}}
\phi (y_1) \cdots \phi (y_n), & \nonumber
\end{eqnarray}
where
$P = (H,{\bf P})$
is four--momentum and we made use of the invariance of the trace
under cyclic permutations of the operators. The essential point is that
the operators $P$ in the vacuum Hilbert space
satisfy the relativistic spectrum
condition, i.e.,
$H \geq |{\bf P}|$.
Hence, disregarding mathematical subtleties, one is led by
inspection of the right hand side of (3) to the conclusion that
this expression can be analytically continued in $z$ into the tube
\begin{equation} \Tb = \{ z \in \CC^4 : | \mbox{Im} \, {\bf z} | <
\mbox{Im} \, z_0 < \beta - | \mbox{Im} \, {\bf z} |  \}.
\end{equation}
This region is considerably larger than the domain in the usual
formulation of the KMS--condition which involves only the time
variable. Moreover, if one proceeds on the right hand of (3)
with the variable $z$
inside of
$\Tb$
to the boundary point
$z = i (\beta, 0)$,
one recovers the KMS--boundary--condition, i.e.,
\begin{eqnarray}
& \langle
\phi (x_1) \cdots \phi (x_m)
\phi (y_1 + i (\beta,0)) \cdots \phi (y_n + i (\beta,0))
\rangle_\beta = & \\
& = \langle \phi (y_1) \cdots \phi (y_n)
\phi (x_1) \cdots \phi (x_m)
\rangle_\beta & \nonumber
\end{eqnarray}
in a somewhat sloppy notation. Any thermal equilibrium state with
these stronger analyticity and boundary value properties is said
to satisfy the {\em relativistic KMS--condition}.

The above heuristic argument can be cast into a rigorous proof under
rather general assumptions \cite{BrBu1}. There exist also variants
where one ends up with slightly weaker results, i.e., a somewhat
smaller domain of analyticity than
$\Tb$.
It has been argued in \cite{BrBu1} that those cases may be of
relevance, e.g., at phase transition points. But we do not enter
into these subtleties here.

The relativistic KMS--condition has been established in simple
models by explicit computations. Moreover, it has been verified in
perturbation theory \cite{St}. It therefore seems to be a
reasonable input in a general discussion of thermal equilibrium
states. In a sense, it replaces the relativistic spectrum condition
in the vacuum sector. Note that the domain
$\Tb$
tends to the forward tube
$\RR^4 + i V_+$
for
$\beta \rightarrow \infty$.
Thus in this limit one recovers the well--known analyticity
properties of vacuum expectation values.

As a first application of the relativistic KMS--condition
in the form stated above let us note
that the correlation functions
$ \langle
\phi (x_1) \cdots \phi (x_{n+1})
\rangle_\beta $
admit in the corresponding set of spacetime variables
$x_2 - x_1, \dots  x_{n+1} - x_n$
an analytic continuation into the union of domains
$(\alpha_1 \Tb) \times \dots \times (\alpha_n \Tb)$
for
$ \alpha_i > 0, i=1, \dots n$,
and
$ \sum_{i=1}^n \alpha_i = 1$.
This follows from standard arguments in the theory of analytic
functions of several complex variables (flat tube theorem). As a
consequence, one can multiply the correlation functions without
running into any ambiguities. This fact allows one for example to
define Wick powers of generalized free fields in thermal states
satisfying the relativistic KMS--condition and
it is also of relevance in perturbation theory.

The relativistic KMS--condition also manifests itself in specific
momentum space properties of the correlation functions. Roughly
speaking, the Fourier transforms
$ \langle \hat{\phi} (p_1 ) \cdots \hat{\phi} (p_{n+1}) \rangle_\beta$
of these functions decrease exponentially when either one of the
four--momenta $p_1 + \dots + p_m$, $1 \leq m \leq n$, tends to
infinity outside of the closed forward lightcone
$\overline{V}_+$.
This feature resembles to some extent the support properties of the
vacuum expectation values in momentum space. But it does not give
rise to analogous relations between momentum space Green
functions.

Let us now turn to the discussion of the consequences of locality.
Infering from the vacuum case, where the interplay between the
relativistic spectrum condition and locality leads to an
enlargement of the domain of analyticity of the correlation
functions (the so--called primitive domains), it seems likely that
similar results also hold in the case of thermal equilibrium states
satisfying the relativistic KMS--condition. One may hope to
establish in this way a rigorous relation between the Euclidean
and Minkowski space formulations of the theory, in analogy to the
Osterwalder--Schrader theorem \cite{OsSch}, cf.\ also \cite{Fr}.
Such a result would be an important step towards the consolidation
of the theory and seems worth some efforts.

Another field of applications of the general formalism is the
derivation of representations of thermal correlation functions
which comply with the condition of locality and the (relativistic)
KMS--condition. In view of the mathematical difficulties which one
faces in this context already in the vacuum case it is clear that
one may only hope to establish such representations for correlation
functions involving a rather small number of fields $(n = 2, 3)$. But
these functions contain already some
very interesting physical information. Moreover, they are important
ingredients both, in the construction and interpretation of the
theory.

Some progress on this problem has been made in \cite{BrBu2} for the
case of the two--point functions. There the condition of locality
leads to strong constraints on the form of these functions which
allow one to apply the techniques of the Jost--Lehmann--Dyson
representation. Let us focus attention here on the case of
spatially homogeneous and isotropic thermal equilibrium states.
Then the two--point function has the form
\begin{equation}
\langle \phi (y) \phi (x) \rangle_\beta =
\Wb ( x-y ),
\end{equation}
where
$\Wb (x)$
is the Fourier transform of a positive, polynomially bounded
density (measure) according to Bochners theorem. As has been shown
in \cite{BrBu2}, the condition of locality (1) implies that the
functions
$\Wb(x)$
can be represented in the form
\begin{equation}
\Wb (x) = \int_0^\infty dm \,
{\cal D}_\beta ( {\bf x}, m ) \,
\Wb^{(0)} ( x,m ).
\end{equation}
Here
$
{\cal D}_\beta ( {\bf x}, m ) $
is a distribution in
$ {\bf x}, m$
which is spherically symmetric in
${\bf x}$,
and
\begin{equation}
\Wb^{(0)} (x,m) = (2 \pi)^{-3} \int d^{\, 4} p \,
\varepsilon (p_0) \delta (p^2 - m^2) (1 - e^{- \beta p_0} )^{-1}
e^{i px}
\end{equation}
is the two--point correlation function of a free field of mass $m$ in
a thermal equilibrium state at inverse temperature
$\beta$.
The precise definition of the expression on the right hand side of
relation (7) requires some care, it is to be understood in the
sense of distributions \cite{BrBu2}. But if the underlying
equilibrium state satisfies the relativistic KMS--condition, the
function
$
{\cal D}_\beta ( {\bf x}, m ) $
is regular in
$ {\bf x} $
and admits an analytic continuation into the domain
$ \{ {\bf z} \in \CC^{\, 3} : | \mbox{Im} \, {\bf z} | < \beta / 2 \}$.
We also note that the positivity of the Fourier transform of
$ \Wb (x)$
imposes further constraints on
$ {\cal D}_\beta ( {\bf x}, m ) $.
The ensuing conditions are rather technical and we therefore omit
them here. But we note that positivity holds if the Fourier
transform
$
\hat{{\cal D}}_\beta ( {\bf p}, m ) $ of
$ {\cal D}_\beta ( {\bf x}, m ) $
is positive.

The representation (7) of thermal two--point functions has the same
general form as the K\"all\'en--Lehmann representation in the vacuum
case. It is a superposition of free correlation functions which are
modulated by amplitude factors
$
{\cal D}_\beta ( {\bf x}, m ) $. In contrast to the vacuum case
these factors depend in general in a non--trivial way on the spatial
variables
${\bf x}$.
They describe the dissipative effects of the thermal background
on the propagation of sharply localized excitations. Appealing
to the intuitive picture that this propagation ought to be damped,
the functions
$
{\cal D}_\beta ( {\bf x}, m ) $
were called damping factors in \cite{BrBu3}. In view of their
physical significance, it is of interest that the damping factors
can be recovered from the correlation functions
$\Wb (x)$, namely that
relation (7) can be inverted. There holds
\begin{equation}
{\cal D}_\beta ( {\bf x}, m ) =
- 2 \pi i {\partial \over \partial m} \int d x_0 \,
J_0 ( m \sqrt{x^2}) \,  \big( \Wb (x) - \Wb (-x) \big),
\end{equation}
where
$J_0$
is the zeroth order Bessel function of first kind. Again, this
expression is well--defined in the sense of distributions.

{}From relation (7) one can derive analogous representations for the
time--ordered, retarded and advanced two--point functions. The
required multiplication with step functions in the time variable
can be handled by standard distribution theoretic methods
\cite{BrBu2}. Similarly, one obtains from relation (7) by analytic
continuation in the time variable the general form of two--point
functions in the imaginary time formalism \cite{Ka}. Let us finally
comment on the formalism of thermofield dynamics \cite{Um}. There
one complements the basic field
$\phi$
by another field
$\tilde \phi$,
called ``tilde'' or ``type two field''.
As is well--known \cite{Oj}, the correlation
functions involving the tilde field
$\tilde \phi$
are completely fixed by those of the original field
$\phi$.
For example, there holds
\begin{eqnarray}
& \langle \tilde \phi (x ) \tilde \phi (y) \rangle_\beta  =
\langle \phi (y) \phi (x) \rangle_\beta & \\
& \langle \tilde \phi (x ) \phi (y) \rangle_\beta  =
\langle \phi (y) \tilde \phi (x) \rangle_\beta =
\langle \phi (y ) \phi (x + i ( \beta /2, 0)) \rangle_\beta, &
\nonumber
\end{eqnarray}
where the last expression is a shorthand for the analytic
continuation of the respective function in the time variable
$x_0$.
Plugging this information into relation (6) one finds that the
representations of the two--point functions involving tilde fields
are obtained by replacing in relation (7) the free field function
$\Wb^{(0)} (x, m)$ by the respective functions
involving the free tilde fields. As before,
one can then proceed to the time--ordered functions etc. Thus
relation (7) provides the desired information on the general form
of thermal two--point functions in all standard approaches to the
theory of thermal equilibrium states in relativistic quantum
field theory.

We conclude this survey with some remarks pertaining to possible
applications of the representation (7).

{\em 1. Sum--rules\/} Assuming that the field
$\phi (x)$
satisfies equal time commutation relations, one concludes from
relation (7) that
\begin{equation}
\int_0^\infty dm \,
{\cal D}_\beta ( 0, m )  = 1
\end{equation}
for all inverse temperatures
$\beta$.
In the case of theories  where the equal time commutation relations
have to be "renormalized" and the integral (11) does not exist, one may
still expect that the differences
$
{\cal D}_{\beta'} ( 0, m )  -
{\cal D}_{\beta''} ( 0, m ) $
integrate to
$0$
for arbitrary
$\beta', \beta'' > 0$.
It seems plausible that this feature is a consequence of the
physically well--motivated assumption that the short distance
singularities of quantum fields (in the sense of the Wilson
expansion) do not depend on the underlying thermal equilibrium
state, cf.\ the principle of local definiteness in \cite{Ha}.
Relation (11) or its generalizations could then be used to
renormalize the fields consistently in all temperature states.

{\em 2. Particle--interpretation\/} It is well--known that
particle--like constituents of a thermal state do not manifest
themselves
by discrete mass shell contributions in the Fourier transforms
of the two--point correlation functions in the presence of
interaction \cite{NaReTh}. But, as it
has been pointed out in \cite{BrBu3},
relations (7) and (9) may help to identify these
constituents. Namely one can associate discrete
($\delta$--function) mass
contributions in
$ {\cal D}_\beta ( {\bf x}, m )$
with stable constituents of the thermal
state. The existence of such contributions is not in conflict with
the results of \cite{NaReTh}. For if
$
{\cal D}_\beta ( {\bf x}, m )$
exhibits a non--trivial
${\bf x}$--dependence, the discrete mass shell singularity in the
Fourier
transform of
$\Wb^{(0)} (x, m)$
is wiped out by convolution with the Fourier--transform of
$
{\cal D}_\beta ( {\bf x}, m )$,
in accord with the heuristic interpretation of these damping
factors.

A simple example \cite{BrBu3}
which nicely illustrates this phenomenon is the
two--point function given by
\begin{equation}
\Wb (x) = \exp{\big( \beta/{2 \lambda} - \sqrt{({\bf x} / \lambda)^2
+ (\beta/{2 \lambda})^2} \big) } \, \Wb^{(0)} (x,m)
\end{equation}
for fixed
$\lambda >0, m > 0$.
It has all the general properties required of a thermal two--point
function in relativistic quantum field theory and satisfies  the
relativistic KMS--condition.

The Fourier transform of this function
can be calculated explicitly. If
$\beta / \lambda$
is large, it decreases off
the mass--shell $p^2 = m^2$ like a Gaussian with variance
proportional to
$1 / \lambda \beta$.
Hence at sufficiently
low temperatures the two--point function describes a
particle--like constituent of the thermal state with an almost
sharp dispersion law. If
$\beta / \lambda$
is small one can extract from the Fourier transform of
$ \Wb (x)$
a contribution which is of Breit--Wigner type, provided the spatial
momentum
$| {\bf p} |$
is large compared to
$\lambda^{-1}$.
This contribution has poles at
$p_0 = \pm E \pm i | {\bf p} | / {\lambda E}$,
where
$E = \sqrt{ {\bf p}^2 + m^2}$.

The latter result can easily be understood if one interprets
$\lambda $ as the
mean free path of a particle in the thermal background. The condition
$| {\bf p} | \gg \lambda^{-1}$
then says that the de Broglie wave length of the particle has to be
small compared to its mean free path. This is clearly a necessary
condition if one wants to identify the constituents of a thermal
state. (Note that the situation is different for
quasi--particles, i.e., collective excitations of a thermal state.)
Because of the pole--structure one is led to ascribe to particles
with velocity
$v = | {\bf p} | / E$
a ``mean lifetime''
$\tau = \lambda / v$.
This result is consistent with the interpretation of
$\lambda$
since within the time
$\tau$
the particle covers the distance
$\lambda$
and therefore is likely to interact with another particle. It then
ceases to exist as a singly--localized excitation of the thermal state
and consequently does not contribute anymore to the two--point
function.

It is a very interesting problem whether the constituents of a
quark--gluon plasma, say, can be described by such discrete
mass--contributions in the decomposition (7) of the pertinent
two--point functions.

{\em 3. Generalized free field approach to perturbation theory\/}
It has been pointed out by Landsman \cite{La} that free field theory
might not be the appropriate starting point for a perturbative
treatment of thermal states since particles in a thermal
environment are ``unstable''. One should take into account this fact
already in lowest order and start from generalized free fields
with suitable weight functions. Although we do not subscribe to the
idea that the constituents of a thermal state are to be
regarded as unstable (their contributions to the two--point
functions are suppressed simply because of statistical reasons,
for they participate with overwhelming probability in collision
processes), the consideration of the dissipative effects of the
thermal background by such an ansatz seems to be meaningful.

The general representation (7) provides a convenient starting point
for this approach since it describes all generalized free field
theories which are admissible according to the general principles:
one simply picks damping factors
$
{\cal D}_\beta ( {\bf x}, m ) $, inserts them into relation
(7) and puts all truncated $n$--point functions with
$n > 2$
equal to
$0$.
An ansatz for the two--point function such as in relation (12)
effectively leads to an infrared--regularized perturbation expansion.
We recall in this context that Wick--powers of generalized
free fields satisfying the relativistic KMS--condition
can be defined without problems. So all necessary ingredients for
a systematic perturbation theory in the sense of Epstein--Glaser
\cite{EpGl} or Steinmann \cite{St} are at hand, but the consistency
of this approach has yet to be established.

Generalized free fields have also been used in self--consistent
computations of thermal states \cite{He}. We expect that the
representation (7) will prove to be useful in these applications
as well.

{\em 4. Goldstone theorem\/}
The general form of the two--point function
is also of relevance in the derivation of Goldstone--type theorems,
which provide information on spectral properties of energy and
momentum in the case of spontaneous symmetry breaking. The
situation is not quite as simple as in the vacuum case, however,
since the damping factors do not need to be measures with respect
to the mass $m$. But the two--point functions of thermal states still
exhibit characteristic momentum space singularities in the
presence of spontaneous symmetry breaking \cite{BrBu4}.

Summing up, the present results show that the model--independent
analysis of the properties of thermal equilibrium states leads to new
and useful insights, even though there are still many open problems
which call for further investigations.
But by combining the efforts in the general
structural analysis with the insights gained in the
investigation of concrete models, one may hope to reach eventually
an understanding of thermal states in relativistic quantum field theory
which is similar to that of the vacuum state. \\[5mm]
\noindent
{\Large\bf Acknowledgements} \\[1mm]
Financial support of this project by PROCOPE and
a travel grant by Deutsche
Forschungsgemeinschaft (DFG)
are gratefully acknowledged.

\end{document}